**Serhii Nazarovets**
http://orcid.org/0000-0002-5067-4498

# Visualization of rank-citation curves for fast detection of h-index anomalies in university metrics

**Abstract.** University rankings, despite facing criticism, continue to maintain their popularity. In the 2023 Scopus Ranking of Ukrainian Universities, certain institutions stood out due to their high h-index, despite modest publication and citation numbers. This phenomenon can be attributed to influential research topics or involvement in international collaborative research. However, these results may also be due to the authors' own efforts to increase the number of citations of their publications in order to improve their h-index. To investigate this, the publications from the top 30 universities in the ranking were analysed, revealing humpback rank-citation curves for two universities. These humpbacks indicate unusual trends in the citation data, especially considering the high percentage of self-citations and FWCI of analysed papers. While quantitative analysis has limitations, the combination of humped rank-citation curves, self-citations, FWCI, and previous research findings raises concerns about the possible causes of these anomalies in the citation data of the analysed universities. The method presented in this paper can aid ranking compilers and citation databases managers in identifying potential instances of citation data anomalies, emphasizing the importance of expert assessment for accurate conclusions.

Despite facing constant criticism, global and national university rankings that utilize a combination of formalised indicators to measure research productivity and impact continue to maintain their popularity worldwide (Lazarev et al., 2017; Gadd, Holmes & Shearer, 2021; Hamann & Ringel, 2023). Government officials use university rankings to make decisions about funding allocations (Ahlers & Christmann-Budian, 2023). University administrators try to compare their own institution's performance with that of other institutions in order to conduct strategic planning at the level of the whole university or individual departments. Students and researchers consider rankings when deciding where to study or work (Brankovic et al., 2023). Furthermore, prior research findings indicate that although the rankings' indicators may not have a direct impact on the allocation of funding, university administrators and staff gradually shift their attention towards these indicators over time (Aagaard, 2015).

This steady popularity of university rankings, both in the national context and internationally, is associated with globalisation, geopolitics and the spread of the audit culture (Shore & Wright, 2015; Brankovic et al., 2023). Ukraine is not exempt from these ranking trends either, as it has more than two hundred universities and potential consumers of their services, mainly applicants and their parents, want a tool to compare universities. Since the state maintenance offered for research and education of the Ukrainian universities depend mainly on the number of students enrolled in the institution, their administrations are also interested in high rankings, which can be used for marketing purposes to attract domestic and foreign applicants (Kovtunets & Londar, 2019; Hladchenko, 2023). Few Ukrainian institutions appear in widely recognised university rankings: only 42 appear in the Times Higher Education (THE) World University Rankings, only 11 in the QS World University Rankings and none in the Academic Ranking of World Universities (ARWU). As a result, national university rankings are more favoured in Ukraine because, unlike international rankings, they cover a wider range of Ukrainian institutions.

One of these common rankings is The Scopus Ranking of Ukrainian Universities[1], which uses only three indicators: the number of publications, the number of citations and the university's h-index. The methodology of this Ukrainian ranking is poor, even compared to the imperfect methodologies of the popular global rankings (Universiteiten van Nederland, 2023). This is because the ranking solely focuses on publication activity indicators, derived from a single bibliographical source, ignoring other vital aspects of university work. The universities in this Ukrainian national ranking are ranked based on the h-index (including self-citations). This indicator aims to provide a harmonious representation of both research productivity and the research impact (Hirsch, 2005). However, the h-index cannot be considered the best measure to represent an institution's scientific impact overall (Waltman & van Eck, 2012). As an alternative to the h-index, scientometricians have proposed the highly cited publications indicator (Waltman & van Eck, 2012) or the π-index (Vinkler, 2009), which do not lead to controversial rankings. Nevertheless, the creators of the Ukrainian ranking did nothing to reduce the likelihood that the h-index ranking could be misused to make incorrect cross-sectoral comparisons (Iglesias & Pecharromán, 2007). Despite the methodology's glaring limitations, this rating continues to be popular in Ukraine. Perhaps the secret of its popularity lies in the fact that the data for this ranking is obtained from an independent source, and therefore, Ukrainian users consider ranking based on this data to be more objective than a national assessment involving of local experts.

In the recently published 2023 Scopus Ranking of Ukrainian Universities (Table 1), I was particularly intrigued by the performance of certain institutions. Despite their relatively modest publication count and citation numbers, these universities demonstrated a high h-index.

The observed phenomenon can be primarily attributed to the influential nature of the research topics addressed within these institutions. The choice of such topics tends to garner higher citation rates due to their relevance and impact in the field. Alternatively, it is possible that these institutions have researchers who actively engage in collaborative efforts with larger research networks, potentially leading to increased citations for their publications (Tahamtan, Safipour Afshar & Ahamdzadeh, 2016). However, given that the ranking position of a university in the aforementioned Ukrainian ranking is primarily determined by the h-index, another plausible explanation could be that the targeted activities of the authors (and university administrators) are aimed at increasing the citations of relevant publications in order to increase the university scientometric indicators.

---

[1] http://osvita.ua/vnz/rating/88976/

**Table 1. TOP-30 Scopus Ranking of Ukrainian Universities, 2023**

|     | University | Papers | Citations | h-index |
| --- | --- | --- | --- | --- |
| 1.  | Taras Shevchenko National University of Kyiv | 24098 | 179925 | 117 |
| 2.  | V. N. Karazin Kharkiv National University | 12916 | 90242 | 85 |
| 3.  | Ivan Franko National University of Lviv | 9300 | 65539 | 74 |
| 4.  | Odesa I. I. Mechnikov National University | 4473 | 32974 | 72 |
| 5.  | National Technical University of Ukraine "Igor Sikorsky Kyiv Polytechnic Institute" | 11670 | 53056 | 70 |
| 6.  | Yuriy Fedkovych Chernivtsi National University | 4624 | 25920 | 69 |
| 7.  | Sumy State University | 4527 | 41373 | 62 |
| 8.  | Lviv Polytechnic National University | 11702 | 56413 | 61 |
| 9.  | Danylo Halytsky Lviv National Medical University | 2079 | 18705 | 60 |
| 10. | Vasyl Stefanyk Precarpathian National University | 1646 | 19474 | 59 |
| 11. | National Technical University Kharkiv Polytechnic Institute | 6076 | 28433 | 57 |
| 12. | Donetsk State Medical University | 1546 | 12415 | 55 |
| 13. | Dnipro State Medical University | 833 | 20217 | 53 |
| 14. | Oles Honchar Dnipro National University | 5009 | 24102 | 52 |
| 15. | Bogomolets National Medical University | 1930 | 11510 | 51 |
| 16. | Ukrainian State Chemical Technology University | 1883 | 13612 | 48 |
| 17. | Kharkiv National Medical University | 2352 | 10399 | 48 |
| 18. | Uzhhorod National University | 3444 | 19405 | 47 |
| 19. | Shupyk National Healthcare University of Ukraine | 1310 | 11498 | 45 |
| 20. | Bohdan Khmelnytsky National University of Cherkasy | 677 | 8519 | 45 |
| 21. | National University of Life and Environmental Sciences of Ukraine | 2956 | 13465 | 42 |
| 22. | National Aerospace University "Kharkiv Aviation Institute" | 2404 | 14220 | 41 |
| 23. | Kharkiv National University of Radio Electronics | 4490 | 16024 | 40 |
| 24. | Kremenchuk Mykhailo Ostrohradskyi National University | 939 | 6205 | 40 |
| 25. | Vasyl' Stus Donetsk National University | 2366 | 9707 | 38 |
| 26. | National University of Kyiv-Mohyla Academy | 1018 | 8562 | 38 |
| 27. | Zaporizhzhia State Medical University | 2978 | 10963 | 37 |
| 28. | Vernadsky Taurida National University | 1047 | 8047 | 36 |
| 29. | Dnipro University of Technology | 1928 | 7767 | 36 |
| 30. | Lesya Ukrainka Volyn National University | 1312 | 9810 | 35 |

To validate this assumption, I selected publications from the top 30 universities in Scopus Ranking of Ukrainian Universities between the years 2003 and 2022. On 6-7 May 2023, I generated rank-citation curves for the publications of each university. In these visualizations,

I plotted the rank of publications along the horizontal axis and the corresponding citation counts on the left axis. Based on the findings of Ye and Rousseau (2010), I obtained smooth ranking curves for most Ukrainian universities in this top 30 (Fig. 1).

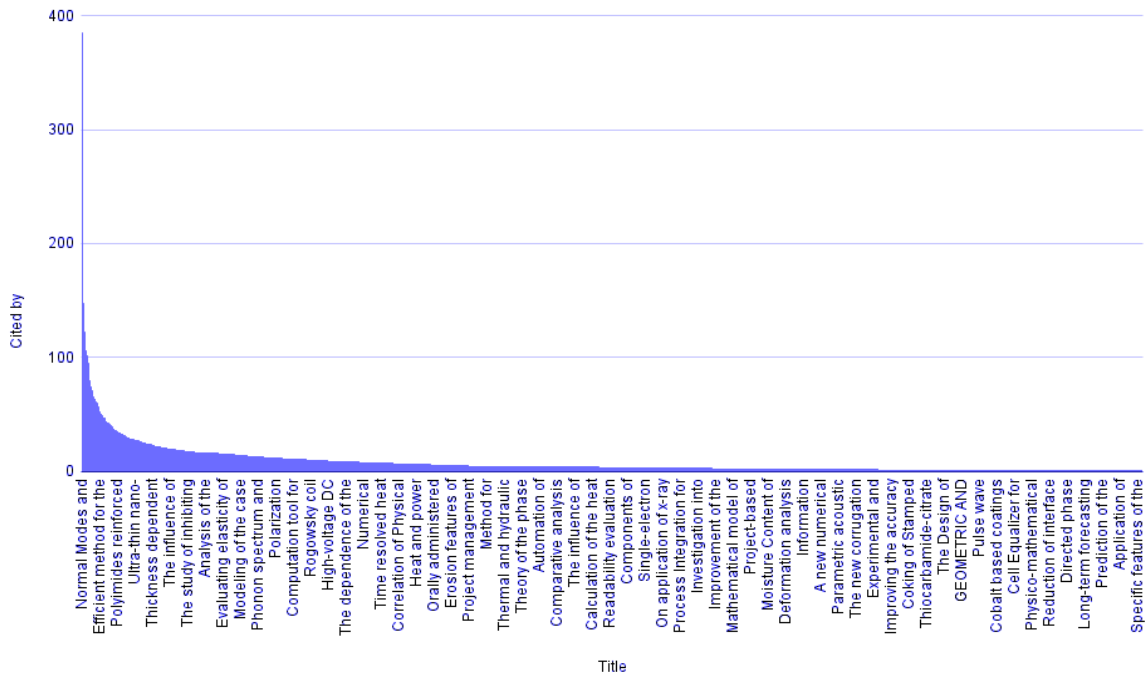

**Fig. 1. Typical rank-citation curve of universities in the top 30 Scopus Ranking of Ukrainian Universities (on the example of the National Technical University Kharkiv Polytechnic Institute)**

However, for two universities from our TOP-30, namely Kremenchuk Mykhailo Ostrohradskyi National University and Dnipro University of Technology, I obtained rank curves with humps (Fig. 2-3).

Could these humpback rank-citation curves indicates an artificial influence on the h-index values of these universities? Our visualizations are very similar to the earlier results of Bartneck & Kokkelmans (2011, 89), who write: "If we now look at the citation index of the unfair author, we notice a humpback around the h-paper... An author with a random self-citation strategy does not have such a humpback. This may come at no surprise, since the humpback is a direct result of self-citing papers close to the h-paper".

I also decided to additionally check which publications belong to these humps and who cited these papers. In the case of the Kremenchuk Mykhailo Ostrohradskyi National University, there are 39 publications responsible for this hump, each with 40 to 45 citations and cited in 667 publications according to Scopus, 356 of which are self-citations (53.4%). For the Dnipro University of Technology, there are 45 publications responsible for the hump, each with 34 to 41 citations, and according to Scopus, they are cited in 721 publications, including 299 self-citations (41.5%).

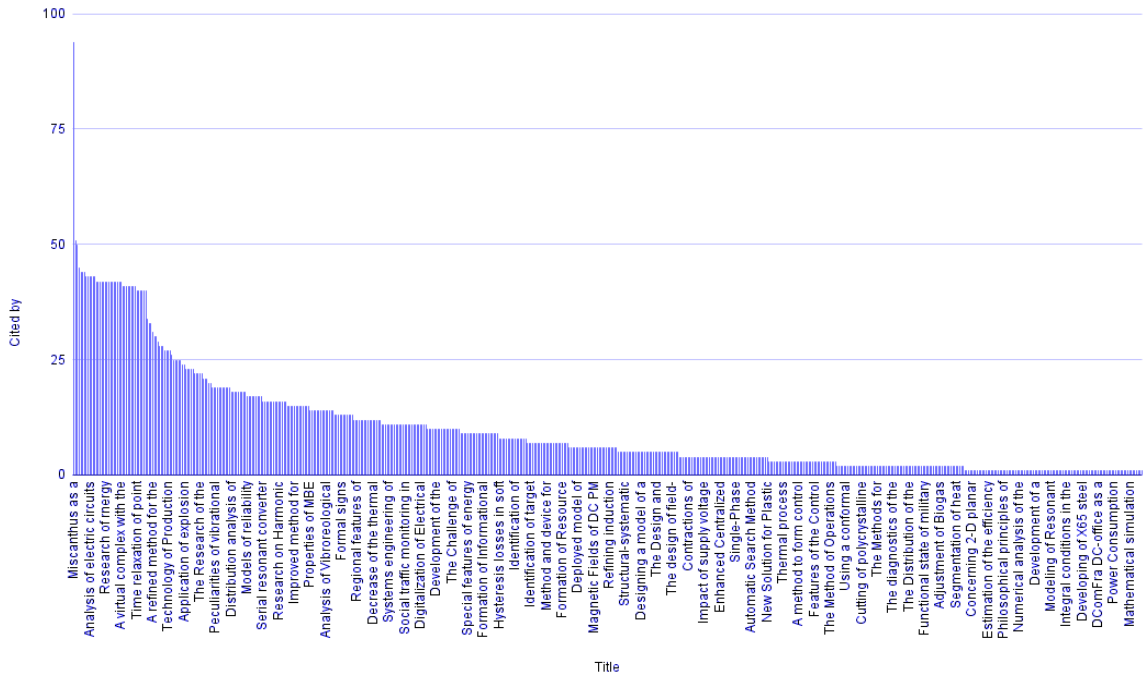

**Fig. 2. Humpback rank-citation curve of the Kremenchuk Mykhailo Ostrohradskyi National University**

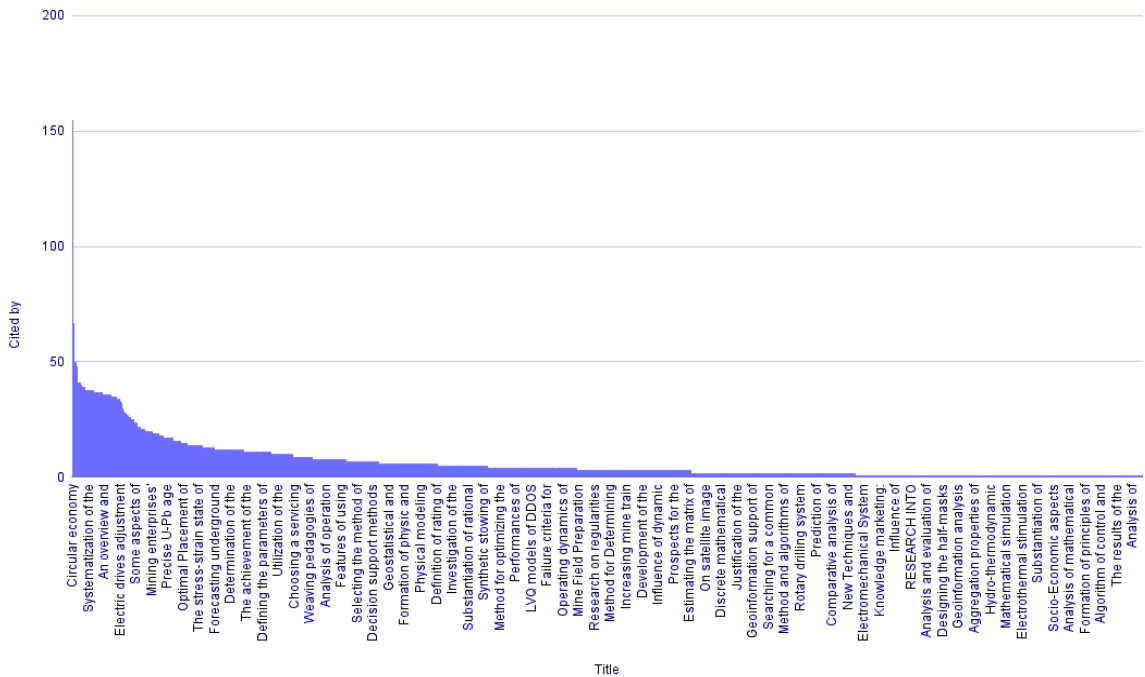

**Fig. 3. Humpback rank-citation curve of the Dnipro University of Technology**

Using SciVal, I also found out the Field-Weighted Citation Impact (FWCI) of these selected publications. For the selected 39 publications of the Kremenchuk Mykhailo Ostrohradskyi National University, the FWCI is 3.35, and for the selected 45 publications of the Dnipro University of Technology, the FWCI is 4.32. That is, the publications I analysed were cited 3 and 4 times better than other publications of the same year and in the same scientific field.

I acknowledge the limitations of quantitative analysis and recognise that the percentages of self-citations and FWCI of publications alone do not provide convincing explanations for the emergence of these unusual patterns of citation data in these two Ukrainian universities. However, when considered alongside the humped rank curves and the findings of Bartneck & Kokkelmans (2011), the performance of these universities raises significant concerns. To draw reliable conclusions about the causes of these anomalies in the citation data, it would be necessary to have experts in the relevant fields of knowledge evaluate the publications belonging to the humps. The approach outlined in this paper may prove useful for compilers of rankings to unearth unusual trends in citation data. As a result, bibliographic database managers can promptly recognise journals and conferences implicated in such practices. Nevertheless, it is vital to stress that these initial suspicions must be cautiously confirmed with the assistance of experts' evaluations. Additionally, it should be noted that the aforementioned ranking uses general citation count of universities as a whole. Instead, taking into account the full-time equivalence (FTE) of different university's units, the peculiarities of the distribution of citations (and self-citations) in different fields would give us a more accurate picture of the research impact of the university's papers, and also about the involvement of specific authors and groups in the appearance of anomalies in the university's citation data.

The main purpose of this short paper is not to draw the community's attention to the unexpected variations in the citation data of peripheral Ukrainian universities, but to look at the root of this problem. It is worth agreeing with (Horton, 2023) that the rankings, in their current form, promote an unhealthy "game of winners and losers" and competition rather than cooperation between universities. Therefore, instead of focusing their efforts on achieving the main mission, such as implementing open science innovations to increase trust in research by improving transparency and accountability (Nazarovets, 2022), the focus of Ukrainian universities is shifting to achieve certain quantitative indicators. In my opinion, ranking systems provide university staff with dubious benchmarks, and this problem is particularly acute in developing countries. The development plans of such universities are somewhat similar to the rituals of the cargo cults from Tanna Island – they attempt to imitate the statistical indicators of the world's leading universities in global rankings (such as achieving a high h-index), without proper consideration of the context, factors and conditions for achieving these indicators – necessary conditions that universities in developing countries most likely do not have.

Today, the global academia is actively reconsidering the significance of university rankings in the realm of scholarly communication (Gadd, 2020; Nassiri-Ansari & McCoy, 2023). Thus, in response to some of the problematic consequences of university rankings, the *More Than Our Rank*[2] initiative has been developed to enable universities to showcase the various ways they serve the world that are not reflected in the rankings. Another important practical step was recently taken by Utrecht University, which deliberately decided not to submit its data in order not to be included in the Times Higher Education (THE) World University Ranking 2024[3]. It is vital to sustain and advance these crucial initiatives. Nonetheless, it is equally imperative to impart knowledge from top-tier universities to developing countries' universities regarding the perils of depending on rankings when mapping out university research.

---

[2] https://inorms.net/more-than-our-rank/
[3] https://www.uu.nl/en/news/why-uu-is-missing-in-the-the-ranking